\documentclass[journal]{IEEEtran}
\usepackage{amsmath}
\usepackage{amssymb}
\usepackage{latexsym}
\usepackage{multirow}
\usepackage{graphics}
\usepackage{mathrsfs}
\usepackage{multicol}
\usepackage{epsfig}
\usepackage{amsmath}
\usepackage[nooneline,flushleft]{caption2}

\newcommand{\beq}{\begin{equation}}
\newcommand{\enq}{\end{equation}}
\newcommand{\ben}{\begin{eqnarray}}
\newcommand{\enn}{\end{eqnarray}}
\newcommand{\bei}{\begin{itemize}}
\newcommand{\eni}{\end{itemize}}

\begin{document}

\title{Increasing Security Degree of Freedom in Multi-user and Multi-eve Systems}

\author{Kun~Xie and Wen~Chen,~\IEEEmembership{Senior Member,~IEEE} and Lili Wei,~\IEEEmembership{Member,~IEEE} \\
\thanks{Copyright (c) 2012 IEEE. Personal use of this material is permitted. However, permission to use this material for any other purposes must be obtained from the IEEE by sending a request to pubs-permissions@ieee.org.}
\thanks{The authors are with the Department of Electronic Engineering,
Shanghai Jiao Tong University, Shanghai, China, 200240 (e-mail:
\{xiekunuestc, wenchen, liliwei@sjtu.edu.cn.)}
\thanks{This work is supported by the National 973 Project \#2012CB316106 and
\#2009CB824904, by NSF China \#60972031 and \#61161130529.}
}

\maketitle

\begin{abstract}
Secure communication in the Multi-user and Multi-eavesdropper (MUME)
scenario is considered in this paper. It has be shown that secrecy
can be improved when the transmitter simultaneously transmits
information-bearing signal to the intended receivers and artificial
noise to confuse the eavesdroppers. Several processing schemes have
been proposed to limit the co-channel interference (CCI). In this
paper, we propose the increasing security degree of freedom (ISDF)
method, which takes idea from the dirty-paper coding (DPC) and ZF
beam-forming. By means of known interference pre-cancelation at the
transmitter, we design each precoder according to the previously
designed precoding matrices, rather than other users' channels,
which in return provides extra freedom for the design of precoders.
Simulations demonstrate that the proposed method achieves the better
performance and relatively low complexity.
\end{abstract}

\begin{IEEEkeywords}
MUME-MIMO, Block Diagonalization,  ZF beam-forming, ISDF, Secrecy
Capacity,
\end{IEEEkeywords}


\section{Introduction}

\IEEEPARstart{T}{he} growing interest in security at the physical
layer of wireless communications has sparked a resurgence of
research in secure communication. In the early works on information
theoretic security, Wyner introduces the wiretap channel model, in
which the eavesdropper's channel is defined to be a degraded version
of the legitimate receiver's channel \cite{1}. It is shown that a
non-zero secrecy capacity can be obtained only if the eavesdropper's
channel is of lower quality than that of the intended recipient.
Csisz\'ar and K\"orner extend this problem to a general non-degraded
channel condition in which a common message is transmitted to the
two receivers and the confidential message to only one of them
\cite{2}. Another main assumption in the aforementioned works is
that the eavesdropper's channel is known at the transmitter
\cite{3,4}, and then the generalized singular value decomposition (GSVD)
\footnote{The generalized singular value decomposition (GSVD) of an
$m\times n$ matrix $\mathbf{A}$ and a $p\times n$ matrix
$\mathbf{B}$ is given by the pair of factorizations
~$\mathbf{A}=\mathbf{U}\sum_1[\mathbf{0,R}]\mathbf{Q}^T$ and
$\mathbf{B}=\mathbf{V}\sum_2[\mathbf{0,R}]\mathbf{Q}^T$, where
$\mathbf{U}$, $\mathbf{V}$ and $\mathbf{Q}$ are orthogonal matrices,
$R$ is an $r\times r$ upper triangular nonsingular matrix, $\sum_1$
and $\sum_2$ are nonnegative diagonal matrice satisfying
$\sum_1^T\sum_1+\sum_2^T\sum_2=\mathbf{I}$.} method can be used to transmit the signal
to the null space of the channel from transmitter to eavesdropper. Clearly,
these assumptions are usually impractical and unreasonable, particularly for passive eavesdroppers. In this paper, we overcome this problem and propose a scheme without using any CSI of eavesdroppers.


In order to achieve secure communication, even when the receiver's
channel is worse than the eavesdropper's channel, or the absence of
eavesdroppers' channel state information (CSI), various
physical-layer techniques have been proposed. One of the most common
techniques is the use of cooperative interference or artificial
noise to confuse the eavesdropper. The cooperative interference
method can be divided into two categories: (i) the trust-friend
model, in which two base stations connected by a high-capacity
backbone such as optical fiber, and one base station can
continuously transmit an interfering signal to secure the uplink
communication for the other base station \cite{5,6}; (ii) the
helper-relay model, where the secrecy level can be increased by
having the cooperative interferer \cite{7} or relay \cite{8} to
send codewords independent to the source message, which can be
canceled at the intended receiver.

Another major techniques for secure communication is the use of multiple antennas.
When multiple antennas are equipped at the transmitter, it is
possible for the transmitter to simultaneously transmit both the information-bearing
signal and artificial noise to achieve secrecy in
a fading environment \cite{9}-\cite{11}, which may replace the role
of the cooperative interference method in \cite{5}-\cite{8}. In the design of secure
communication with artificial noise, the transmit power allocation
between the information signal and the artificial noise is an
important issue, which has not been discussed in \cite{9,10,11}.
A suboptimal power allocation strategy is considered in \cite{12},
which aims to meet an ideal signal-to-interference-and-noise ratio
(SINR) at the intended receiver to satisfy a quality of service
requirement. The secure communication with artificial noise is also
discussed in \cite{11}, in which the closed-form expression of achievable
rate and the optimal power allocation has been obtained, however only
single-receiver and single-antenna at receiver was considered.

Most of the previous papers focus on the single-user systems.
However most practical communication systems have more than one
user and the eavesdroppers may not appear alone as well, and they
may choose to cooperate or not \cite{11}. In addition, each terminal may
be equipped with multiple antennas, which is representative, for example,
of downlink transmission in LTE systems and wireless local area networks.
This is the so called Multi-User and Multi-Eve (MUME) MIMO systems, which have
been seldom investigated before. In this paper, we will focus on investigating the MUME systems.

It's also worth noting that the achievable
secrecy rate of MUME systems is different from that of single-user and single-antenna
systems studied before, which must make sure that any legitimate user will not
be wiretapped by any eavesdropper. The authors in \cite{14} put forward an MUME model
in which all other users are viewed as potential eavesdroppers by the targeted user.
They also give a definition of the achievable secrecy rate of multiuser
wiretap model in terms of \emph{secrecy sum rate}. The authors in \cite{15} give another
definition of achievable secrecy rate in Gaussian
MIMO multi-receiver Wiretap channel, which is named secrecy capacity region.
Besides, the authors in \cite{16,17} considers the compound wire-tap
channel, which is based on the classical wire-tap channel with channel from the source to the destination and the channel from the source to the wiretapper taking a number of states respectively. It can be viewed as the multicast wire-tap channel with multiple
destinations and multiple wire-tappers with the same massage transmitted
to different destinations, which is slightly different from the broadcast wire-tap
channels in this paper. They also give another significant definition of the
achievable secrecy rate in terms of \emph{absolute secrecy rate}, which idea is to take the
security of the poorest-performance receive-wiretap pair into consideration. If the poorest-performance pair can meet the
quality of service requirement, then all other pairs can do. Therefore this definition of achievable secrecy rate may be more reasonable and constructive in the practical secure communication systems. The authors in \cite{17,18} discussed the realization of the
achievable secrecy rate of multiple users (multiple eavesdroppers) with artificial noise separately, in which, however, the system model is compound wiretap channel but broadcast MUME wiretap channel~\cite{xiekun1}.


Since the transmitter needs to transmit different message to different
receivers in the broadcast MUME wiretap model, there must be considerable
co-channel interference (CCI) in the system. In order to limit the CCI from
the signals transmitted to other users and mask receivers' own message
signal simultaneously, two practical linear transmission schemes were often used
in the early works: (i) the SVD method discussed in \cite{9,11}, which conducts an SVD decomposition on each user's channel matrix to get a maximum channel gain for their
own message but can not suppress the interference from other user's message;
(ii) the ZF beamforming method \cite{19} and its promotion----the BD
method \cite{20,21}, in which all the information is transmitted in
the null space of all other receivers' channels. The SVD method and
ZF beamforming method are simple, but of little ideal performance.
While the BD method is of somewhat ideal performance but more complicated than
the formers.

In view of the drawbacks of the previous schemes, we propose an
alternative approach, which takes idea from dirty-paper coding (DPC)
\cite{22,23} and ZF beam-forming \cite{19}. It can directly increase
the degree of freedom when designing the transmission precoders, which in
return make obvious improvement not only at the achievable secrecy rate, but
also at the antenna constraints at transmitter compared with the BD method.
What's more, in our proposed scheme, we choose to map the artificial
noise into the null space of co-precoder matrix instead of the
null space of legitimate receivers' channels in the existing schemes,
which may offer extra improvement on the secrecy rate
in the low SNR region. The performance will be further improved when
the water-filling (WF) method is used. Besides, the power allocation
between the information signal and the artificial noise are also discussed.

In this paper, ${(\cdot)}^H$ and $tr(\cdot)$ denote the Hermitian
transpose and trace of a matrix. $E(\cdot)$ denotes expectation, and
$\mathbf{I}$ denotes an identity matrix. $I(\cdot,\cdot)$ denotes
mutual information. $[x]^+$ = max$\{0,x\}$.

\section{System Model}

In this paper, we consider the broadcast MUME wiretap model as shown in
Fig. 1, in which there is one transmitter named Alice, $J$
legitimate users named Bobs and $K$ passive eavesdroppers named Eves.
Alice tries to send independent messages to all the legitimate receivers
while keeping each of the eavesdropper ignorant of all the messages. All of
the terminals are equipped with multiple antennas. $N_{Bj}$ antennas
are equipped at the $j$-th Bob, $N_{Ek}$ antennas at the $k$-th Eve,
and $N_A$ antennas at the single Alice. This scenario is representative,
for example, of downlink transmission in the LTE systems and wireless local
area networks.

Let the transmit signal ${\bf X}=\sum_{j=1}^J{{\bf U}_j} + {\bf V}$,
where ${\bf U}_{j}$ is the information bearing signal vector for
user $j$, and ${\bf V}$ is the artificial noise signal vector to
interference Eves. Then the received signals at Bobs and Eves are
respectively:
\begin{figure}
  \captionstyle{flushleft}
  \onelinecaptionstrue
  \centering
  \includegraphics[width=3.6in]{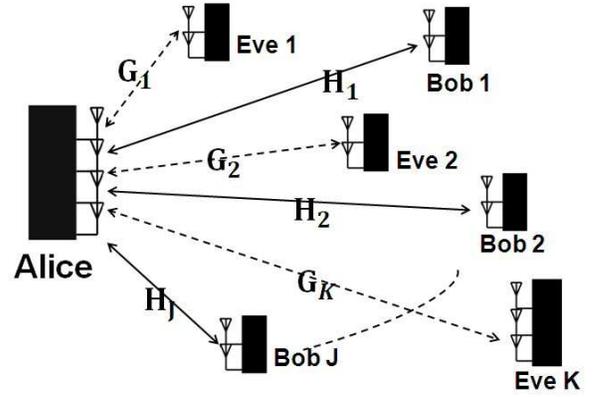}\label{F1 system model }
  \caption{The MUME-MIMO wiretap system model.}
\end{figure}
\begin{equation}
\begin{aligned}
&\text{Bob}~ j:~~{\bf Y}_{j}={\bf H}_{j}{\bf X}+{\bf N}^B_{j},~~\text{for}~ j = 1, . . . ,J ,\\
&\text{Eve}~ k:~~{\bf Z}_{k}={\bf G}_{k}{\bf X}+{\bf N}^E_{k},~~\text{for}~ k = 1, . . . ,K ,\label{L1 equation 1}
\end{aligned}
\end{equation}
\\
where ${\bf H}_j$ is the $N_{Bj} \times N_A$ channel matrix between
the transmitter and Bob $j$,
%
${\bf G}_k$ is the $ N_{Ek}\times N_A$ channel matrix between the
transmitter and eavesdropper $k$,
%
%
$\mathbf{N}^B_{j}$ and  $\mathbf{N}^E_{k}$ are respectively the
additive white Gaussian noise vectors observed at the $j$-th Bob and
$k$-th Eve, which covariance matrices satisfy $E[\mathbf{N}^B_{j}
{\mathbf{N}^B_{j}}^H ]=\sigma_{Bj}^2 \mathbf{I}$, and
$E[\mathbf{N}^E_{k} {\mathbf{N}^E_{k}}^H ]=\sigma_{Ek}^2 \mathbf{I}$
respectively.

We assume that the channel matrix $\mathbf{H}_j$ and ${\bf G}_k$ are
block-fading, whose entries are complex Gaussian variables with
zero-mean and unit-variance. We also assume that perfect channel
state information (CSI) of the receiver, i.e., the channel matrices
$\mathbf{H}_j$, $j = 1,2,~\ldots~,J$, are available at Alice, e.g.,
either through reverse channel estimation in time-division-duplex
(TDD) or feedback in frequency-division-duplex (FDD). While the
channel matrices $\mathbf{G}_k$, $k = 1,2,~\ldots~,K$, are
unavailable at Alice due to the passive nature of eavesdroppers.

Our objective is to transmit different secret message to the
corresponding Bobs. We try to reduce the CCI from the
others, and make sure that the underlying Eves can not wiretap any
communication between Alice and Bobs. In the following, we provide
lower and upper bounds on the achievable secrecy rate of the generalized
broadcast MUME wire-tap channel.

Let $R^B_j$ denote the mutual information rate between Alice and Bob $j$, and $R^E_k$ denote that between Alice and Eve $k$ for Bob $j$. Then
\begin{equation}
\begin{aligned}
&R^B_{j}=\max[I(U_j;Y_j)], ~~\text{for} ~j = 1, . . . ,J,\\
&R^E_{jk}=\max[I(U_j;Z_k)], ~~\text{for} ~k = 1, . . . ,K. \label{L2
equation 2}
\end{aligned}
\end{equation}
In the sequel, the achievable secrecy rate of the receive-wiretap pair $(j,k)$ (for Bob $j$ and Eve $k$) can be denoted by \cite{24}
\begin{equation}
\begin{aligned}
R_{jk}={[R^B_{j}-R^E_{jk}]}^+.
\end{aligned}
\end{equation} \label{L3 equation 3}

The achievable secrecy rate of MUME wiretap model is usually noted by
secrecy sum rate ($R_{ssr}$) \cite{14} or secrecy rate region
\cite{15}. However, in the practical broadcast MUME wireless communication
systems, the massages transmitted vary with different users, we should take
each user into consideration, once it is chosen according to some criterion.
Therefore we must make sure that any user's communication can not be
wiretapped by any eavesdropper. Hence the secrecy rate of the system
is determined neither by the best transmission pair nor the total rate gap
between Bobs and Eves, but by the poorest-performance transmission pair in
reality. Then we propose an alternative definition of the secrecy rate
for MUME wire-tap channel, which is called absolute secrecy rate ($R_{asr}$).
Obviously, the absolute secrecy rate actually is just the lower
bound of the achievable secrecy rate, and can be given by
\begin{equation}
\begin{aligned}
R_{asr}&=\min_{j,k}\{R_{jk}\}=\min_{j,k}\{{[R^B_{j}-R^E_{jk}]}^+\}={[\min_{j}\{R^B_{j}\}-\max_{k}\{R^E_{jk}\}]}^+\\
&={[\min_{j}\max_{P_{U_j}}I(Y_j;U_j)-\max_{k}\max_{P_{U_j}}I(Z_k;U_j))]}^+,
\end{aligned}
\end{equation}\label{L4 Definition 4}
where $P_{U_j}$ is an input distribution.
As for the secrecy sum rate $(R_{ssr})$, we have
\begin{equation}
\begin{aligned}
R_{ssr}&=\sum_j\min_{k}\{R_{jk}\}=\sum_{j}{[R^B_{j}-\max_k\{R^E_{k}\}]}^+\\
&=\sum_j{[\max_{P_{U_j}}I(Y_j;U_j)-\max_{k}\max_{P_{U_j}}I(Z_k;U_j))]}^+.
\end{aligned}
\end{equation}\label{L5 Definition 5}

\section{the Design of Precoders in MUME-MIMO Network Based on ISDF}

At Alice, the data for each user is processed before transmission. Then it is launched into the MIMO channel with the random artificial noise. Let $\mathbf{W}_j$ be an $N_{A}\times d_j$ linear precoder, $\mathbf{u}_j$ be a $d_j\times 1$ symbol vector for Bob $j$, and $d_j$ be the number of parallel data symbols transmitted simultaneously for Bob $j$ \cite{22} satisfying $1\leq d_j \leq N_{Bj}$. Let $\mathbf{V}$ be the artificial noise signal vector, $\mathbf{W}$ be the transmission preprocessing matrix, and $\mathbf{v}$ be the symbol vector. They are both used for the artificial noise. Then the transmission signal is
\begin{equation}
\begin{aligned}
&\mathbf{X}=\sum_{j=1}^J \mathbf{U}_j+\mathbf{V}=\sum_{j=1}^J\mathbf{W}_j\mathbf{u}_j+\mathbf{W}\mathbf{v}.\\
\end{aligned}\label{L6 equation 6}
\end{equation}
The received signal at Bobs and Eves are respectively
\begin{equation}
\begin{aligned}
&\mathbf{Y}_j=\mathbf{H}_j\sum_{\ell=1}^J \mathbf{W}_\ell\mathbf{u}_\ell+\mathbf{H}_j\mathbf{Wv}+\mathbf{N}^B_j, ~j = 1,~ \ldots ~,J ,\\
&\mathbf{Z}_k=\mathbf{G}_k\sum_{\ell=1}^J
\mathbf{W}_\ell\mathbf{u}_\ell+\mathbf{G}_k\mathbf{Wv}+\mathbf{N}^E_k,
~k = 1,~ \ldots ~,K . \label{L7 equation 7}
\end{aligned}
\end{equation}

The emphasis of this paper is to design the precoders
$\mathbf{W}_\ell$  for $\ell=1,2,~\ldots~,J$ and $\mathbf{W}$. In
the SVD method all the data is transmitted in their own channel
image space, so that each Bob can get the maximum channel gain for
the corresponding message. Another one is the block diagonalization
(BD) method, in which all the data is transmitted in the null space
of all other Bobs, which is the promotion of the ZF beamforming
method and can reduce the interference from the other users' message
signal. Here, we propose an alternative approach----the increasing
security degrees of freedom (ISDF) method, which is based on
dirty-paper coding (DPC) \cite{22,23} and ZF beam-forming \cite{19}.
In this paper, a single data stream is to be sent to each receiver
when $d_j=1,~\forall j$, and multiple data streams are sent when
$d_j>1,~\forall j$. Note that, only a maximum of $N_{Bj}$ streams
can be transmitted simultaneously for user $j$, else the message
will not be decoded. Criterion of judging the design is whether the
secrecy rate is sufficiently good under the given power
constraints, which will be discussed in detail with the design of
precoders in the following part.

In MUME scenarios, several co-channel Bobs with multiple antennas
aim to communicate with Alice in the same frequency and time slots.
In this case, it is necessary to design transmission scheme that is
able to suppress the CCI at Bobs. In multi-user wireless security
communications with artificial noise, the precoding matrix is
usually designed in the null space of channel matrix $\mathbf{H}_j$.
We may called it precoding selection space as well. Obviously, the smaller
the rank of matrix $\mathbf{H}_j$ is, the larger the dimension of its
corresponding precoding selection space will be. In return, the design
of the corresponding precoder has more freedom and the secrecy performance
will be better. Therefore we define the dimension of the precoding
selection space as the security degrees of freedom (SDF).


\subsection{Design of Precoders for Bobs Based on ISDF}

In the existing schemes e.g., the SVD method, ZF beamforming method
and BD method, the precoder is designed based on their corresponding
user's own channel matrix or the other users' channel matrices,
which means the SDF will be largely limited by the rank of their
corresponding channel matrices and the performance will be
inevitably affected. To solve this problem, here we propose a new
method, which are designed based on the previously designed
precoders instead. Because the new method can directly increase the
SDF when designing each precoder, we just name it as the increasing
security degree of freedom (ISDF) method. This method is similar to
the idea of DPC method \cite{23} in some sense. For example, Alice
first picks a precoder for Bob~1 and then chooses a precoder for Bob
2 with full (noncausal) knowledge of the precoder for Bob 1.
Therefore, Bob 1 does not see the signal intended for Bob 2 as
interference. Similarly, the precoder for Bob 3 is chosen such that
Bob 1 and Bob 2 do not see the signals intended for Bob 3 as
interference. This process continues for all Bobs. Bob J
subsequently sees the signals intended for all other users as
interference, Bob 2 sees the signals intended for Bob 1 as
interference, etc.

Suppose that the $J$ Bobs has been sorted as Bob 1, Bob 2, $\ldots$,
Bob $J$ according to some criterion, which will be discussed in the
simulation section. We first design $\mathbf{W}_1$ for Bob 1 without
loss of generality.
\begin{equation}
\begin{aligned}
\mathbf{W}_1 \propto \max~d_1~\text{eigenvectors
~of}~(\mathbf{H}_1^H \mathbf{H}_1 ),
\end{aligned}\label{L8 equation 8}
\end{equation}
where Eq. $(\ref{L8 equation 8})$ means that $\mathbf{W}_1$ is composed
by the $d_1$ eigenvectors corresponding to the largest $d_1$
eigenvalues of $\mathbf{H}_1^H \mathbf{H}_1$, $1 \leq d_{1} \leq
N_{B1}$.

Then the following precoders are designed to satisfy an basic
condition that each of them must be located in the null space of all
previously designed precoders.

\begin{equation}
\left\{{
\begin{array}{ll}
\mathbf{W}_2\subset \ker(\mathbf{W}_1),\\
\mathbf{W}_3\subset \ker(\mathbf{W}_1) \bigcap \ker(\mathbf{W}_2),\\
~\vdots\\
\mathbf{W}_j\subset \underset{i=1,i<j}{\bigcap}\ker(\mathbf{W}_i),\\
~\vdots\\
\mathbf{W}_J\subset
\overset{J-1}{\underset{i=1}{\bigcap}}\ker(\mathbf{W}_i),
\end{array}}\right. \label{L9 equation 9}
\end{equation}
where $\ker(\cdot)$ denotes the null space (the kernel) of some
matrix, and $\cap$ represents the intersection of subspaces.
Here, we define the co-designed-precoders matrix for the Bob $j$ as
\begin{equation}
\mathbf{\widetilde{W}}_j={[\mathbf{W}_1,\mathbf{W}_2,~\ldots~,\mathbf{W}_{j-1}]}.
\label{L10 equation 10}
\end{equation}
From Eq. $(\ref{L9 equation 9})$, we know that the design of each
precoder should satisfy
$\mathbf{W}_j^T\mathbf{\widetilde{W}}_j=[\textbf{0},\cdots,\textbf{0}]$.

Let $L_j$ be the dimension of $\ker(\widetilde{\mathbf{W}}_j)$. Then
the precoder $\mathbf{W_j}$ can be composed by
\begin{equation}
\mathbf{W}_j=\mathbf{T}_{null,j}\mathbf{T}_{stream,j},  \label{L11
equation 11}
 \end{equation}
where $\mathbf{T}_{null,j}$ is used to suppress the interference,
which is an $N_A \times L_j$ matrix. $\mathbf{T}_{stream,j}$ is an
$L_j\times d_j$ matrix used for streams selection, which can make
better use of the space resource and therefore improve the capacity.

Note that the precoding matrix $\mathbf{W}_j$ should be a nonzero
matrix, otherwise, no signal is transmitted. To guarantee the
existence of a nonzero precoding matrix, a sufficient condition is
that the number of the transmit antennas is larger than the previous
$j-1$ users' total data streams, i.e.,
\begin{equation}
\begin{aligned}
N_A>\max_{j=1,2,\ldots,J}~~\sum_{i=1}^{J-1} d_i.
\end{aligned}
\end{equation} \label{L12 equation 12 }

Under this sufficient condition, let
$\{\mathbf{t}_j^1,\mathbf{t}_j^2,~\ldots~,\mathbf{t}_j^{L_j} \}$ be
an orthnormal basis of the subspace $ker(\widetilde{\mathbf{W}}_j)$.
Then the kernel space is spanned by the generator
$\mathbf{T}_j^{(0)}=[\mathbf{t}_j^1,\mathbf{t}_j^2,~\ldots~,\mathbf{t}_j^{L_j}]$.
Then Eq. $(\ref{L9 equation 9})$ can be rewritten as
\begin{equation}
\left\{{
\begin{array}{ll}
\mathbf{W}_2\subset \mbox{span} \mathbf{T}_2^{(0)}=\ker(\mathbf{W}_1)=\ker({\mathbf{\widetilde{W}}}_2),\\
\mathbf{W}_3\subset \mbox{span}\mathbf{T}_3^{(0)}=\ker(\mathbf{W}_1) \bigcap \ker(\mathbf{W}_2)=\ker({\mathbf{\widetilde{W}}}_3),\\
~\vdots\\
\mathbf{W}_j\subset \mbox{span}\mathbf{T}_j^{(0)}=\underset{i=1,i<j}{\bigcap}\ker(\mathbf{W}_i)=\ker({\mathbf{\widetilde{W}}}_j),\\
~\vdots\\
\mathbf{W}_J\subset \mbox{span}\mathbf{T}_J^{(0)}=\overset{J-1}{\underset{i=1}{\bigcap}}\ker(\mathbf{W}_J)=\ker({\mathbf{\widetilde{W}}}_J),\\
\end{array}}\right.
\end{equation}\label{L13 equation 13}
where the generator matrix $\mathbf{T}_j^{(0)}$ can be computed
through singular value decomposition (SVD)~\cite{25} as
\begin{equation}
\begin{aligned}
{\mathbf{\widetilde{W}}}_j=
\left[
\begin{array}{ccc}
\mathbf{T}_j^{(1)}  & \mathbf{T}_j^{(0)}
\end{array}
\right]
\left[
\begin{array}{ccc}
\Sigma_W & \mathbf{0} \\
\mathbf{0 }  & \mathbf{0 }
\end{array}
\right]
\left[
\begin{array}{ccc}
\mathbf{R}_j^{(1)} \\
\mathbf{R}_j^{(0)}
\end{array}
\right],
\end{aligned}\label{L14 equation 14}
\end{equation}
where $\Sigma_W$ is a diagonal matrix whose diagonal entries are in
descending order. Then we can get
$\mathbf{T}_{null,j}=\mathbf{T}_j^{(0)}$, which is an $N_A \times
L_j$ matrix.

As for $\mathbf{T}_{stream,j}$ whose role is to linearly combine the
$L_j$ orthnomral basis to compose a precoder with $d_j$ columns, it
therefore can make better use of the space resource. The design of
$\mathbf{T}_{stream,j}$ can be achieved by applying SVD to the equivalent
channel matrix $\widetilde{\mathbf{H}}_j=\mathbf{H}_j \mathbf{T}_{null,j} $,
\begin{equation}
\begin{aligned}
\widetilde{\mathbf{H}}_j^T= \left[
\begin{array}{ccc}
\widetilde{\mathbf{T}}_j^{(1)}  & \widetilde{\mathbf{T}}_j^{(0)}
\end{array}
\right]
\left[
\begin{array}{ccc}
\Sigma_H & \mathbf{0} \\
\mathbf{0 }  & \mathbf{0 }
\end{array}
\right]
\left[
\begin{array}{ccc}
\widetilde{\mathbf{R}}_j^{(1)} \\
\widetilde{\mathbf{R}}_j^{(0)}
\end{array}
\right],
\end{aligned}\label{L15 equation 15}
\end{equation}
where $\Sigma_H$ is a diagonal matrix whose diagonal entries are in
descending order. Then we get
$\mathbf{T}_{stream,j}=\widetilde{\mathbf{T}}_j^{(1)}$, which is an
$L_j \times d_j$ matrix. Then Eq. $(\ref{L11 equation 11})$ can be
rewritten as:
\begin{equation}\label{L16 equation 16}
\begin{aligned}
\mathbf{W}_j=\mathbf{T}_j^{(0)} \widetilde{\mathbf{T}}_j^{(1)}.
\end{aligned}
\end{equation}

The dimension of $\mathbf{W}_j$ is $d_j\leq L_{j}$. If $d_j=1$,
there is a single data stream sent to Bob $j$, and $\mathbf{T}_{stream,j}$
contains the singular vector corresponding to the largest singular
value, i.e., the data stream is transmitted through the equivalent
channel with the largest singular value. So does for $1<d_j<L_j$. If
$d_j=L_j$, the data streams will be transmitted through all the
sub-channels with non-zero singular value. In order to simplify the
analysis, we assume that the power are uniformly allocated for the
message of user $j$. The secrecy rate can be further increased
if  Water-Filling (WF) method is used, which will be discussed in
section IV.

Obviously, we can get the SDF for each method as following:
\begin{equation}
\begin{aligned}
&\mathbf{SDF}_{j,ISDF}= ~\text{the dimension of}~\cap_{\ell=1}^{j-1}\ker({\mathbf{{W}}}_\ell),\\
&\mathbf{SDF}_{j,BD}= ~\text{the dimension of}~\cap_{\ell\neq j}\ker({\mathbf{{H}}}_\ell),\\
&\mathbf{SDF}_{j,ZF}= ~\text{the dimension of}~\cap_{\ell\neq j}\ker({\mathbf{{H}}}_\ell),\\
&\mathbf{SDF}_{j,SVD}= ~\text{the dimension of}~\ker({\mathbf{H}}_j).
\end{aligned}\label{L17 equation 17}
\end{equation}
Since the dimension of $\ker({\mathbf{{W}}}_\ell)$ is usually
greater than that of $\cap_{\ell\neq j}\ker({\mathbf{{H}}}_\ell)$, and the
intersection operation in ISDF scheme takes less terms than that of
BD and ZF schems, we have the SDF of the ISDF scheme is usually
greater than those of the BD and ZF-beamforming schemes. But the SDF
of the ISDF scheme may not be greater than that of the SVD scheme.
Since the SVD method can not well cancel the CCI, the secrecy
performance of ISDF scheme outperforms that of the SVD scheme, which
will be verified by the simulation results.

\subsection{Precoder Design for Eves Based on ISDF}

Since the CSI of all receivers (except for the eavesdroppers) is
available at the transmitter, in order to guarantee that it does not
impact the desired receivers, the artificial noise is often mapped
into the subspace orthogonal to the effective downlink co-channel
matrix $\hat{\mathbf{H}}$ [11][17], where
\begin{equation}
\begin{aligned}
\hat{\mathbf{H}} ={[\mathbf{H}_1^T,\mathbf{H}_2^T, \ldots
,\mathbf{H}_J^T]}.
\end{aligned}\label{L18 equation 18}
\end{equation}
Then we can get the precoder $\mathbf{W}\subset ker(\hat{\mathbf{H}})$,
i.e., the kernel of $\hat{\mathbf{H}}$. Note that the precoding
matrix $\mathbf{W}$ should also be a nonzero matrix. To guarantee
the existence of a nonzero power of artificial noise, a sufficient
condition is that the number of the transmit antennas is larger than
the rank of matrix $\hat{\mathbf{H}}$. Because the practical channel
matrix is usually assumed to be full-rank, $N_A$ must satisfies
$N_A>\sum_{j=1}^J N_{Bj}$, which is a very tight constraint.

However, we actually don't have to make $\textbf{W}$ orthogonal to
each user's channel matrix $\mathbf{H}_j$. We can achieve the goal
by transmitting the artificial noise into the null space of all
users' precoder matrices instead. Define the effective co-precoder
matrix as
\begin{equation}
\begin{aligned}
\hat{\mathbf{W}} ={[\mathbf{W}_1,\mathbf{W}_2, \ldots ,\mathbf{W}_J]}.\label{L19 equation 19}
\end{aligned}
\end{equation}
To transmit the artificial noise more effectively, it may be mapped
into the subspace orthogonal to the effective co-precoder matrix
$\hat{\mathbf{W}}$. Then we can get $\mathbf{W}\subset
ker(\hat{\mathbf{W}})$, i.e., $\mathbf{W}$ lies in the null
space of $\hat{\mathbf{W}}$. Because inequality $d_{j}\leq N_{Bj}$ is always valid,
the rank of $\hat{\mathbf{W}}$ is usually smaller than that of
$\hat{\mathbf{H}}$. Therefore, we have more freedom to transmit the artificial
noise, and the constraint on $N_A$ can therefore be relaxed as
$N_A>\sum_{i=1}^J d_{j}$.

To distinguish the two schemes of transmitting artificial noise, we
category them as the ISDF1 scheme and the ISDF2 scheme. If the noise
is mapped into the subspace orthogonal to the effective downlink
co-channel matrix $\hat{\mathbf{H}} $, it is called ISDF1; if If the
noise is mapped into the subspace orthogonal to the effective
co-precoder matrix $\hat{\mathbf{W}}$, it is called ISDF2. In
the simulation section, we will compare the two schemes.

\subsection{The analysis of complexity}

In this section, we will make a discussion on the computational complexity of
the proposed approach versus the other threes methods. As introduced in section II,
The SVD method and ZF beamforming method are simple, but of little ideal performance.
The BD method get somewhat better performance on secrecy rate, but its computational
complexity becomes higher compared with the former.

Then, the emphasis is the comparison of complexity between the proposed ISDF method
and the BD method. Essentially, the main difference between the ISDF and BD method lies on
solving the precoding selection matrix (PSM). In the ISDF method, the PSM is
obtained by implementing an SVD decomposition on the previously designed precoders
$\mathbf{\widetilde{W}}_j$, which is a $N_A \times \sum^{j-1}_{n=1}d_n$ matrix.
The complexity of this SVD decomposition is $O((\sum^{j-1}_{n=1}d_n)^3)$.
While the PSM in the BD method is obtained by an SVD decomposition on
all others' channel matrixes $\mathbf{\widetilde{H}}_j$, where
$\mathbf{\widetilde{H}}_j = {[\mathbf{H}^T_1,\mathbf{H}^T_2, \ldots , \mathbf{H}^T_{j-1}, \mathbf{H}^T_{j+1}, \ldots ,\mathbf{H}^T_J]}$, which is a $N_A \times \sum_{n\neq j}N_{Bn}$ matrix. And the complexity of this SVD decomposition is $O((\sum_{n\neq j}N_{Bn})^3)$,
which is higher than the former. Since both of the two methods have $J$ precoders to design,
there are $J$ PSMs need to be solven accordingly. Therefore, the complexity of
the ISDF method is $O(\sum^J_{j=1}(\sum^{j-1}_{n=1}d_n)^3)$,
on the other hand, that of the BD method is $O(\sum_{j=1}^J(\sum_{n\neq j}N_{Bn})^3)$.
Obviously, the complexity of the our proposed approach is lower.

\section{The secrecy rate of MUME-MIMO system}

In this section, we will analyze the secrecy rate of ISDF1 and
ISDF2. Suppose that the variance of the transmit symbol vector
$\mathbf{u}_j$ is ${\sigma_{u_j}^2}$, and the complex Gaussian
random elements of $\mathbf{v}$ are i.i.d whose variance is
${\sigma_v^2}$. It is assumed that Alice has a total amount of
transmit power budget $P$. Due to the normalization of the noise
variance at Bob, we can also refer to $P$ as the transmission SNR.
One important parameter should be designed is the power ratio,
denoted by $\rho_j~~(0< \rho_j <1)$, allocated for the user $j$'s
information transmission. We define the power ratio for transmitting
artificial noise as $\alpha$ ($0< \alpha <1$). Let
\begin{equation}\label{L20 equation 20}
\mathbf{Q}_j\triangleq E(\mathbf{u}_j \mathbf{u}_j^H ),\quad
\mathbf{Q}_v\triangleq E(\mathbf{v} \mathbf{v}^H ).
\end{equation}
Then we have
\begin{equation}\label{L21 equation 21}
tr({\mathbf{Q}_j})=P_j=\rho_j P, \quad tr({\mathbf{Q}_v})=\alpha P,
\end{equation}
and
\begin{equation}\label{L22 equation 22}
P=\sum_{j=1}^J \rho_j P+ \alpha P =\sum_{j=1}^J d_j \sigma_{u_j}^2 +
\left(N_A-\sum_{j=1}^J d_{j}\right) \sigma_v^2,
\end{equation}
in which, we have used the following facts
\begin{equation}
\begin{aligned}
&\alpha=1-\sum_{j=1}^J \rho_j,\\
&N_A \geq \sum_{j=1}^J d_{j}+1,\\
&\sigma_{u_j}^2=\frac{P_j}{d_j}=\frac{\rho_j P}{d_j},\\
&\sigma_{v}^2=\frac{(1-\sum_{j=1}^J \rho_j)P}{N_A-\sum_{j=1}^J
d_{j}}.
\end{aligned}\label{L23 equation 23}
\end{equation}

In order to analyze the secrecy rate concisely, Eq. $(\ref{L7
equation 7})$ can be rewritten as:
\begin{equation}
\begin{aligned}
\mathbf{Y}_j&=\mathbf{H}_j\sum_{i=1}^J \mathbf{W}_i\mathbf{u}_i+\mathbf{H}_j\mathbf{Wv}+\mathbf{N}^B_j\\
&=\mathbf{H}_j\mathbf{W}_j\mathbf{u}_j+\mathbf{H}_j\sum_{i=1,i\neq j}^J\mathbf{W}_i\mathbf{u}_i+\mathbf{H}_j\mathbf{Wv} +\mathbf{N}^B_j\\
&=\widehat{\mathbf{H}}_{jj}\mathbf{u}_j+\sum_{i=1,i\neq j}^J\widehat{\mathbf{H}}_{ji} +\widehat{\mathbf{H}}_{j}\mathbf{v}+\mathbf{N}^B_j\\
\mathbf{Z}_k&=\mathbf{G}_k\sum_{\ell=1}^K \mathbf{W}_\ell\mathbf{u}_\ell+\mathbf{G}_k\mathbf{Wv}+\mathbf{N}^E_k~~~~~~~~\\
&=\mathbf{G}_k\mathbf{W}_\ell\mathbf{u}_\ell+\mathbf{G}_k\sum_{\ell=1,\ell\neq j}^J\mathbf{W}_\ell\mathbf{u}_\ell+\mathbf{G}_k\mathbf{Wv} +\mathbf{N}^E_k\\
&=\widehat{\mathbf{G}}_{kk}\mathbf{u}_k+\sum_{\ell=1,\ell\neq
j}^J\widehat{\mathbf{G}}_{k\ell}+\widehat{\mathbf{G}}_{k}\mathbf{v}+\mathbf{N}^E_k
\end{aligned}\label{L24 equation 24}
\end{equation}
where we have defined
\begin{eqnarray}
\widehat{\mathbf{H}}_{ji}\triangleq \mathbf{H }_j
\mathbf{W}_i,\quad\widehat{\mathbf{H}}_{j}\triangleq \mathbf{H }_j \label{L25 equation 25}
\mathbf{W},\\
\widehat{\mathbf{G}}_{k\ell}\triangleq \mathbf{G}_k
\mathbf{W}_\ell,\quad\widehat{\mathbf{G}}_{k}\triangleq \mathbf{G
}_k \label{L26 equation 26} \mathbf{W},
\end{eqnarray}
for $j,i,\ell=1,2,~\ldots~,J,~k=1,2,~\ldots~,K$.

The secrecy rate is the maximum transmission rate at which the
intended receiver can decode the data with arbitrarily small error,
which is bounded by the difference in the capacity between Alice and
Bob and that between Alice and Eve \cite{2}. In the following part,
the secrecy rate will be given in terms of secrecy sum rate
\cite{14} and absolute secrecy rate, where the secrecy sum rate
is noted by $R_{ssr}$ and the absolute secrecy rate is noted by
$R_{asr}$.

\subsection{The secrecy rate of ISDF1 }

As in \cite{11}, we can normalize the distance of each Bob to  make
the variance of the elements of $\mathbf{H}_j$ equal to unity
without loss of generality, and the noise vector $\mathbf{N}^B_{j}$
is of unit variance. Since the artificial noise is transmitted in
the null space of all legitimate users' matrixes, it will be nulled
in any user's received signal. Then the capacity between Alice and
Bob $j$ is
\begin{equation}
\begin{aligned}
R^B_{j}&=E_{\widehat{\mathbf{H}}}\left\{ {\log}_2 \left|
~\mathbf{I}+{\sigma}_{uj}^2\widehat{\mathbf{H}}_{jj}\widehat{\mathbf{H}}^H_{jj}\bigg(\mathbf{I}+\sum_{i
\neq j}^J
{\sigma}_{ui}^2\widehat{\mathbf{H}}_{ji}\widehat{\mathbf{H}}^H_{ji}\bigg)^{-1}
\right|
\right\}\\
&=E_{\widehat{\mathbf{H}}}\left\{ {\log}_2 \left|
~\mathbf{I}+\frac{\rho_j
P}{d_j}\widehat{\mathbf{H}}_{jj}\widehat{\mathbf{H}}^H_{jj}\bigg(\mathbf{I}+\sum_{i
\neq j}^J \frac{\rho_i P}{d_i}
\widehat{\mathbf{H}}_{ji}\widehat{\mathbf{H}}^H_{ji}\bigg)^{-1}
\label{L27 equation 27} \right| \right\},
\end{aligned}
\end{equation}
where we used the fact $\widehat{\mathbf{H}}_{j}=0$.

Next, we study the capacity between Alice and the multiple colluding
or non-concluding Eves. When multiple Eves are allocated at
different places, the noise at each Eve may be different. In
addition, the receiver noise levels at Eves may not be known by
Alice or Bobs. To guarantee secure communication, it is therefore
reasonable to consider the worst-case scenario where the noises at
Eves are arbitrarily small. Note that this approach has been also
taken in \cite{9} and \cite{11}. In this case, the noiseless
eavesdropper assumption gives an upper bound on the rate between
Alice's message for the user $j$ and the eavesdropper $k$ as
\begin{equation}
\begin{aligned}
R^E_{jk}=&E_{ \widehat{\mathbf{H}},\mathbf{G}_k}  \bigg\{ {\log}_2
\bigg|
\mathbf{I}+\sigma_{uj}^2\widehat{\mathbf{G}}_{kj}{\widehat{\mathbf{G}}_{kj}}^H
 \Big(\sum_{\ell=1,\ell\neq j}^J
{\sigma_{u\ell}^2\widehat{\mathbf{G}}_{k\ell}{\widehat{\mathbf{G}}_{k\ell}}^H}\\
&{\quad+\sigma_{v}^2\widehat{\mathbf{G}}_{k}{\widehat{\mathbf{G}}_{k}}^H\Big)}^{-1}
\bigg|
\bigg\}\\
=&E_{ \widehat{\mathbf{H}},\mathbf{G}_k}  \bigg\{ {\log}_2 \bigg|
\mathbf{I}+\frac{\rho_j
P}{d_j}\widehat{\mathbf{G}}_{kj}{\widehat{\mathbf{G}}_{kj}}^H{\bigg(\sum_{\ell=1,\ell\neq
j}^J \frac{\rho_\ell
P}{d_\ell}\widehat{\mathbf{G}}_{k\ell}{\widehat{\mathbf{G}}_{k\ell}}^H}\\
&{\quad+\frac{\alpha
P}{N_A-\sum_{i=1}^J
N_{Bi}}\widehat{\mathbf{G}}_{k}{\widehat{\mathbf{G}}_{k}}^H\bigg)}^{-1}
\label{L28 equation 28} \bigg| \bigg\}.
\end{aligned}
\end{equation}

After deriving the expressions of $R^B_j$ and $R^E_{kj}$ , the
ergodic secrecy rate can now be obtained as $R_{jk} =
[R^B_j-R^E_{jk}]^+$.

\subsubsection{Secrecy Sum-Rate }

As proposed in \cite{14}, the secrecy rate for Bob $j$ is

\begin{displaymath}
R^j_{se}=\underset{1\leq k\leq K,}{\min}\,\{R_{jk} \}.
\end{displaymath}
Then the secrecy sum rate is
\begin{equation}
\begin{aligned}
R_{ssr}=&\sum_{j=1}^J R^j_{se}=\sum_{j=1}^J\underset{1\leq k\leq K,}{\min}\,\{R_{jk} \}.  \\
=&\sum_{j=1}^J \underset{k}{\min}~E_{ \widehat{\mathbf{H}},\mathbf{G}_k}
\Bigg\{\Bigg[{\log}_2 \bigg| \mathbf{I}+\frac{\rho_j
P}{d_j}\widehat{\mathbf{H}}_{jj}{\widehat{\mathbf{H}}_{jj}}^H
\\&\Big(\mathbf{I}+\sum_{i \neq j}^J \frac{\rho_i P}{d_i}
\widehat{\mathbf{H}}_{ji}\widehat{\mathbf{H}}^H_{ji}\Big)^{-1}
\bigg| \\
&\quad- {\log}_2 \bigg|
\mathbf{I}+\frac{\rho_j
P}{d_j}\widehat{\mathbf{G}}_{kj}{\widehat{\mathbf{G}}_{kj}}^H{\bigg(\sum_{\ell=1,\ell\neq
j}^J \frac{\rho_\ell
P}{d_\ell}\widehat{\mathbf{G}}_{k\ell}{\widehat{\mathbf{G}}_{k\ell}}^H}\\
&{\quad\quad+\frac{\alpha
P}{N_A-\sum_{i=1}^J
N_{Bi}}\widehat{\mathbf{G}}_{k}{\widehat{\mathbf{G}}_{k}}^H\bigg)}^{-1}
\bigg|\Bigg]^+
\Bigg\}.
\end{aligned} \label{L29 equation 29}
\end{equation}

\subsubsection{Absolute secrecy rate }

As we have discussed, for the broadcast MUME-MIMO wiretap system, the
absolute secrecy capacity is the lower bound of the ergodic secrecy rate,
which is given by
\begin{equation}
\begin{aligned}
R_{asr}=&\underset{1\leq j\leq J,1\leq k\leq K}{\min}\,\{R_{jk}
\}=\underset{j,k}{\min}E_{ \widehat{\mathbf{H}},\mathbf{G}_k} \Bigg\{
\Bigg[\\&
{\log}_2 \bigg| \mathbf{I}+\frac{\rho_j
P}{d_j}\widehat{\mathbf{H}}_{jj}{\widehat{\mathbf{H}}_{jj}}^H
\Big(\mathbf{I}+\sum_{i \neq j}^J \frac{\rho_i P}{d_i}
\widehat{\mathbf{H}}_{ji}\widehat{\mathbf{H}}^H_{ji}\Big)^{-1}
\bigg| \\
&\quad- {\log}_2 \bigg|
\mathbf{I}+\frac{\rho_j
P}{d_j}\widehat{\mathbf{G}}_{kj}{\widehat{\mathbf{G}}_{kj}}^H
{\bigg(\sum_{\ell=1,\ell\neq j}^J \frac{\rho_\ell
P}{d_\ell}\widehat{\mathbf{G}}_{k\ell}{\widehat{\mathbf{G}}_{k\ell}}^H}\\
&\quad\quad{+\frac{\alpha
P}{N_A-\sum_{i=1}^J
N_{Bi}}\widehat{\mathbf{G}}_{k}{\widehat{\mathbf{G}}_{k}}^H\bigg)}^{-1}
\bigg|\Bigg]^+
\Bigg\}.
\end{aligned} \label{L30 equation 30}
\end{equation}

\subsection{The secrecy rate of ISDF2 }
In the ISDF2 method, the rate between Alice and Bob $j$ is
\begin{equation}
\begin{aligned}
R^B_{j}&=E_{\widehat{\mathbf{H}}}\bigg\{ {\log}_2 \bigg|
~\mathbf{I}+{\sigma}_{uj}^2\widehat{\mathbf{H}}_{jj}\widehat{\mathbf{H}}^H_{jj}\bigg(\mathbf{I}+\sum_{i
\neq j}^J
{\sigma}_{ui}^2\widehat{\mathbf{H}}_{ji}\widehat{\mathbf{H}}^H_{ji}\\&
\quad+
{\sigma}_{v}^2\widehat{\mathbf{H}}_{j}\widehat{\mathbf{H}}^H_{j}\bigg)^{-1}
\bigg|
\bigg\}\\
&=E_{\widehat{\mathbf{H}}}\bigg\{ {\log}_2 \bigg|
~\mathbf{I}+\frac{\rho_j
P}{d_j}\widehat{\mathbf{H}}_{jj}\widehat{\mathbf{H}}^H_{jj}\bigg(\mathbf{I}+\sum_{i
\neq j}^J \frac{\rho_i P}{d_i}
\widehat{\mathbf{H}}_{ji}\widehat{\mathbf{H}}^H_{ji}\\&
\quad+ \frac{\alpha
P}{N_A-\sum_{i=1}^J
d_{i}}\widehat{\mathbf{H}}_{j}\widehat{\mathbf{H}}^H_{j}\bigg)^{-1}
\bigg| \bigg\}.
\end{aligned}
\end{equation}\label{L31 equation 31}
Similar to Eq. $(\ref{L28 equation 28})$, the rate between Alice's message
for the user $j$ and the eavesdropper $k$ can be rewritten as:
\begin{equation}
\begin{aligned}
R^E_{kj}=&E_{ \widehat{\mathbf{H}},\mathbf{G}_k}  \bigg\{ {\log}_2
\bigg|
\mathbf{I}+\sigma_{uj}^2\widehat{\mathbf{G}}_{kj}{\widehat{\mathbf{G}}_{kj}}^H
\Big(\sum_{\ell=1,\ell\neq j}^J
{\sigma_{u\ell}^2\widehat{\mathbf{G}}_{k\ell}{\widehat{\mathbf{G}}_{k\ell}}^H}\\
&\quad{+\sigma_{v}^2\widehat{\mathbf{G}}_{k}{\widehat{\mathbf{G}}_{k}}^H\Big)}^{-1}
\bigg|
\bigg\}\\
=&E_{ \widehat{\mathbf{H}},\mathbf{G}_k}  \bigg\{ {\log}_2 \bigg|
\mathbf{I}+\frac{\rho_j
P}{d_j}\widehat{\mathbf{G}}_{kj}{\widehat{\mathbf{G}}_{kj}}^H
{\bigg(\sum_{\ell=1,\ell\neq j}^J \frac{\rho_\ell
P}{d_\ell}\widehat{\mathbf{G}}_{k\ell}{\widehat{\mathbf{G}}_{k\ell}}^H}\\
&\quad{+\frac{\alpha
P}{N_A-\sum_{i=1}^J
d_{i}}\widehat{\mathbf{G}}_{k}{\widehat{\mathbf{G}}_{k}}^H\bigg)}^{-1}
\bigg| \bigg\}.
\end{aligned}\label{L32 equation 32}
\end{equation}

\subsubsection{Secrecy Sum-Rate }

Accordingly, the secrecy sum-rate is
\begin{equation}
\begin{aligned}
R_{ssr}=&\sum_{j=1}^J R^j_{se}=\sum_{j=1}^J\underset{1\leq k\leq K,}{\min}\,\{R_{jk} \}.  \\
=&\sum_{j=1}^J \underset{k}{\min}~E_{ \widehat{\mathbf{H}},\mathbf{G}_k}
\Bigg\{ \Bigg[{\log}_2 \bigg| \mathbf{I}+\frac{\rho_j
P}{d_j}\widehat{\mathbf{H}}_{jj}{\widehat{\mathbf{H}}_{jj}}^H\\&
\Big(\mathbf{I}+\sum_{i \neq j}^J \frac{\rho_i P}{d_i}
\widehat{\mathbf{H}}_{ji}\widehat{\mathbf{H}}^H_{ji}+\frac{\alpha
P}{N_A-\sum_{i=1}^J
d_{i}}\widehat{\mathbf{G}}_{k}{\widehat{\mathbf{G}}_{k}}^H\Big)^{-1}
\bigg|\\
&- {\log}_2 \bigg|
\mathbf{I}+\frac{\rho_j
P}{d_j}\widehat{\mathbf{G}}_{kj}{\widehat{\mathbf{G}}_{kj}}^H
{\bigg(\sum_{\ell=1,\ell\neq j}^J \frac{\rho_\ell
P}{d_\ell}\widehat{\mathbf{G}}_{k\ell}{\widehat{\mathbf{G}}_{k\ell}}^H}\\&{+\frac{\alpha
P}{N_A-\sum_{i=1}^J
d_{i}}\widehat{\mathbf{G}}_{k}{\widehat{\mathbf{G}}_{k}}^H\bigg)}^{-1}
\bigg|\Bigg]^+
\Bigg\}.
\end{aligned}  \label{L33 equation 33}
\end{equation}

\subsubsection{Absolute secrecy rate}

we can also get the lower bound of ergodic secrecy rate for ISDF2 method,
\begin{equation}
\begin{aligned}
R_{asr}=&\underset{1\leq j\leq J,1\leq k\leq K}{\min}E_{ \widehat{\mathbf{H}},\mathbf{G}_k} \Bigg\{
\Bigg[{\log}_2 \bigg| \mathbf{I}+\frac{\rho_j
P}{d_j}\widehat{\mathbf{H}}_{jj}{\widehat{\mathbf{H}}_{jj}}^H\\&\Big(\mathbf{I}+\sum_{i
\neq j}^J \frac{\rho_i P}{d_i}
\widehat{\mathbf{H}}_{ji}\widehat{\mathbf{H}}^H_{ji}+\frac{\alpha
P}{N_A-\sum_{i=1}^J
d_{i}}\widehat{\mathbf{G}}_{k}{\widehat{\mathbf{G}}_{k}}^H\Big)^{-1}
\bigg|\\
&- {\log}_2 \bigg|
\mathbf{I}+\frac{\rho_j
P}{d_j}\widehat{\mathbf{G}}_{kj}{\widehat{\mathbf{G}}_{kj}}^H{\bigg(\sum_{\ell=1,\ell\neq
j}^J \frac{\rho_\ell
P}{d_\ell}\widehat{\mathbf{G}}_{k\ell}{\widehat{\mathbf{G}}_{k\ell}}^H}\\&{+\frac{\alpha
P}{N_A-\sum_{i=1}^J
d_{i}}\widehat{\mathbf{G}}_{k}{\widehat{\mathbf{G}}_{k}}^H\bigg)}^{-1}
\bigg|\Bigg]^+
\Bigg\}.
\end{aligned}\label{L34 equation 34}
\end{equation}

\subsection{The secrecy rate of ISDF1 with Water Filling method }

This section focuses on improving the secrecy rate with water-filling
method based on ISDF1 method introduced in section III. As mentioned
in section III-A, when the users have multiple antennas and $d_j$ ($d_j > 1$)
data streams are transmitted to user $j$ simultaneously, the water-filling
method may be employed together with ISDF1 to further improved the
performance of achievable secrecy rate. Our goal is to implement
single-user water-filling on each legitimate user to maximize their rate
under the given power ratio constraint.

The single-user optimization problem has a well-known water-filling
solution. The water-filling algorithm takes advantage of the problem
structure by decomposing the channel into orthogonal modes, which
greatly reduces the optimization complexity. Here, without loss of generality, we just take user $j$ for example with
the constraint of the given total power ($\rho_j P$). As derived in Eq. $(\ref{L24 equation 24})$, the signal received at user $j$
can be written as:
\begin{equation}
\begin{aligned}
\mathbf{Y}_j&= \widehat{\mathbf{H}}_{jj}\mathbf{u}_j+\mathbf{K}_j
\end{aligned}\label{L35 equation 35}
\end{equation}
where $\mathbf{K}_j = \sum_{i=1,i\neq j}^J \widehat{\mathbf{H}}_{ji}u_i +\widehat{\mathbf{H}}_{j}\mathbf{v}+\mathbf{N}^B_j$ is the interference towards
user $j$ from the message signal transmitted to other users. We first implement an Whiting processing on the interference vector $\mathbf{K}_j$ in Eq. $(\ref{L35 equation 35})$, before conducting the water-filling algorithm. Let $M=E(K_jK_j^H)$, which unitary decomposition is $M=E\Lambda E^H$. The we can whiten $K_j$ by
\begin{equation}
\begin{aligned}
\Lambda^{-1/2}E^H\mathbf{Y}_j&= \Lambda^{-1/2}E^H\widehat{\mathbf{H}}_{jj}\mathbf{u}_j+\Lambda^{-1/2}E^HK_j.
\end{aligned}\label{L36 equation 36}
\end{equation}
Then the corresponding mutual information rate $R^B_j$ is

\begin{equation}
\begin{aligned}
R^B_{j}&=E_{\widehat{\mathbf{H}}}\bigg\{ {\log}_2 \bigg|
~\mathbf{I}+{\sigma}_{uj}^2\Lambda^{-1/2}E^H\widehat{\mathbf{H}}_{jj}\widehat{\mathbf{H}}^H_{jj}E\Lambda^{-1/2}
\bigg|
\bigg\}.
\end{aligned}\label{L37 equation 37}
\end{equation}

Suppose that $R_{jn}^B$ is the rate from the $n$-th subcarrier to Bob $j$, and $h_{jn}$ the corresponding effective sub-channel after whitening and orthgononalizing. Since only $d_j$ streams are transmitted to user $j$, the optimal problem on the
dynamic power allocation aiming to maximize the rate $R^B_{j}$
between transmitter and user $j$ can be expressed as following:

\begin{equation}
\begin{aligned}
&~~~~~R^*_j = \max \sum ^{d_j}_{n=1} {\log}_2 \big(1+P_{jn}h_{jn}\big),\\
&s.t.~~\sum ^{d_j}_{n=1}P_{jn}\leq P_j = \rho_j P, ~\text{and}~ P_{jn}>0, ~\forall n,
\end{aligned}\label{L38 equation 38}
\end{equation}
where $P_{jn}$ denotes the power allocated for the $n$-th subcarrier of user $j$.
Based on Eq. $(\ref{L38 equation 38})$, we use the classical Lagrange algorithm to
construct an Lagrange function,

\begin{equation}
\begin{aligned}
&L = \sum ^{d_j}_{n=1} {\log}_2 \left(1+P_{jn}h_{jn}\right) -
\lambda \left(\sum ^{d_j}_{n=1}P_{jn} - \rho_j P\right).
\end{aligned}\label{L39 equation 39}
\end{equation}
Let
\begin{displaymath}
\frac{\partial L}{\partial P_{jn}}=\frac{1}{\ln 2}\frac{h_{jn}}{1+h_{jn}P_{jn}}-\lambda =0,
\end{displaymath}
and $\beta=\lambda \ln 2$. Then we can obtain
$$
\beta  = \frac{h_{jn}}{1+P_{jn}h_{jn}}, \quad P_{jn} = \left[\frac{1}{\beta}-\frac{1}{h_{jn}}\right]^+. \label{L41 equation 41}
$$

Then solving the optimal problem depends on the computation of $\beta$ and $P_{jn}$. The paper \cite{26} put forward a fast iterative algorithm, which give the preliminary value of $\beta$ and its update method. That is
\begin{align}
&\beta_0 = \frac{1}{d_j}\Big(\big|P_j + \sum ^{d_j}_{n=1}\frac{1}{h_{jn}} \big|\Big) \label{L42 equation 42}, \\
&\beta_\ell = \beta_{\ell-1} + \frac{1}{d_j}\Big(\big|P_j + \sum ^{d_j}_{n=1} P_{jn} \big|\Big). \label{L43 equation 43}
\end{align}

In this paper, we will employ this fast iterative algorithm to perform simulations.

\section{Simulation Results}
In this section, we will carry out some simulations to show the
achievable secrecy rate. In all simulations, the entries of all
channel matrices are assumed to be independent, zero-mean Gaussian
random variables with unit variance. All results are based on an
average of $1000$ independent trials. The background noise power is
the same for all Bobs with a variance $\mathbf{I}$. To guarantee the
secure communication, it is therefore reasonable to consider the
worst case scenario, where the noises variance at Eves are
arbitrarily small (approaching zero). The desired rate for Bobs and
Eves will be measured by the ergodic capacity rather than the outage
capacity.

\subsection{Secrecy Rate and Power Efficiency}
Fig. 2, Fig. 3 and Fig.~4, are corresponding to the cases $d_j=1$,
which exhibits the comparison of secrecy rate and information
power ratio for information signal among the $5$ methods. Fig.~2 and
Fig.~3 demonstrate that that the ISDF1 method offers the best
performance compared with the other methods, in term of secrecy sum
rate and absolute secrecy rate respectively. From Fig.~2 and Fig.~3, we can
see that ISDF2 performs best in the low SNR region but performs worse in the high SNR region.
This is because that at the high SNR region, the artificial noise is not well canceled at receive terminal by using the null space of the co-precoder matrix to design $\textbf{W}$. Since ISF1 uses the nulll space of the co-channel matrix to desing $\textbf{W}$, which can well cancel the artificial noise. From Fig.~4, we can see that both ISDF1 and ISDF2 methods have less power efficiency (information power over total power) than the ZF Beamforming
method. While in secure communications, the secrecy rate is the major concern.
Therefore the ISDF scheme provides a good candidate for secure communications.

\begin{figure}
  \captionstyle{flushleft}
  \onelinecaptionstrue
  \centering
  \includegraphics[width=3.6in]{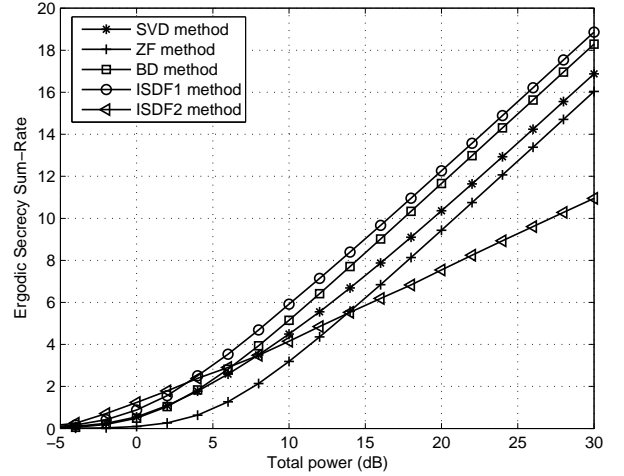}
  \caption{Comparison of secrecy sum rate for the five methods when $J=3$, $K=2$, $d_j=1$, $N_{Bj}=3$, $N_{Ek}=4$.}
\end{figure}


\begin{figure}
  \captionstyle{flushleft}
  \onelinecaptionstrue
  \centering
  \includegraphics[width=3.6in]{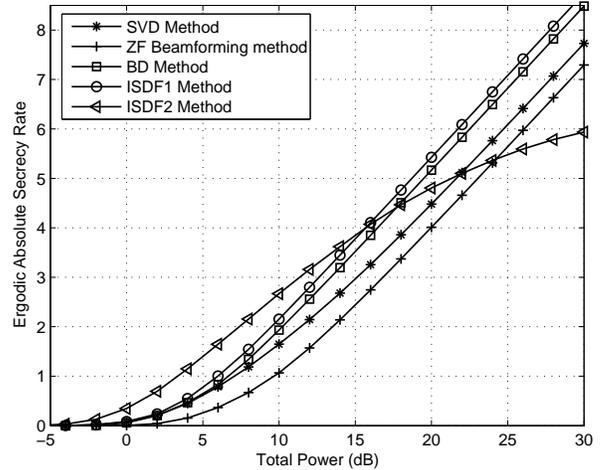}
  \caption{Comparison of absolute secrecy rate for the five methods when $J=3$, $K=2$, $d_j=1$, $N_A=10$, $N_{Bj}=3$, $N_{Ek}=4$.}
\end{figure}

\begin{figure}
  \captionstyle{flushleft}
  \onelinecaptionstrue
  \centering
  \includegraphics[width=3.6in]{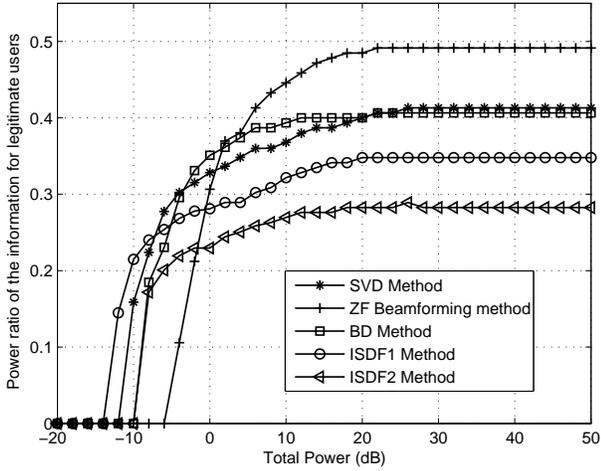}
  \caption{Comparison of information power ratio for the five methods when $J=3$, $K=2$, $d_j=1$, $N_A=10$, $N_{Bj}=3$, $N_{Ek}=4$.}
\end{figure}


\begin{figure}
  \captionstyle{flushleft}
  \onelinecaptionstrue
  \centering
  \includegraphics[width=3.6in]{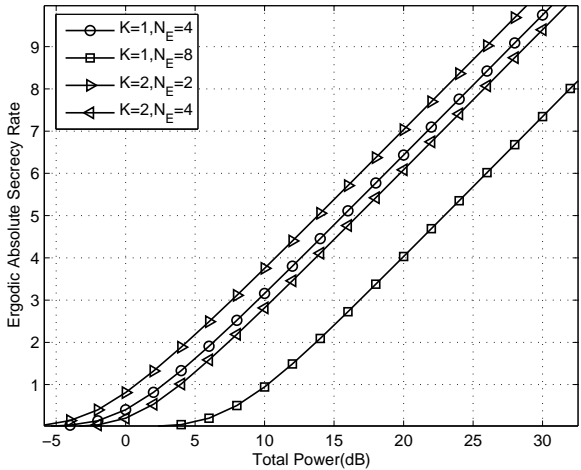}
\caption{The absolute secrecy rate of ISDF1 for Eves' colluding
($K=1$) and non-colluding ($K=2$) scenarios when $J=3$, $d_j=1$, $N_A=10$,
$N_{Bj}=3$. }
\end{figure}

\begin{figure}
  \captionstyle{flushleft}
  \onelinecaptionstrue
  \centering
  \includegraphics[width=3.6in]{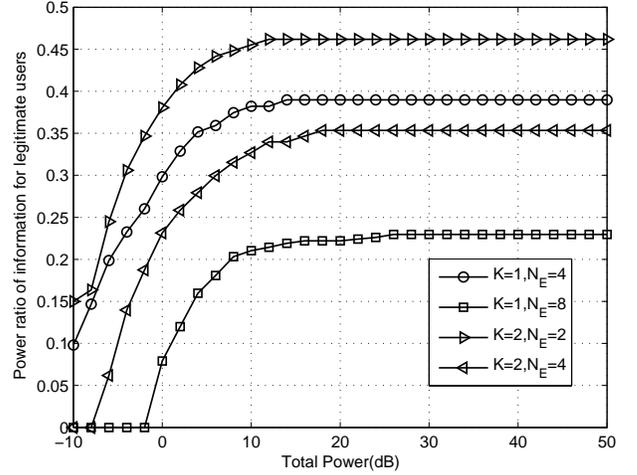}
\caption{The information power ratio of ISDF1 for Eves' colluding
($K=1$) and non-colluding ($K=2$) scenarios when $J=3$, $d_j=1$, $N_A=10$,
$N_{Bj}=3$.}
\end{figure}


\begin{figure}
  \captionstyle{flushleft}
  \onelinecaptionstrue
  \centering
  \includegraphics[width=3.6in]{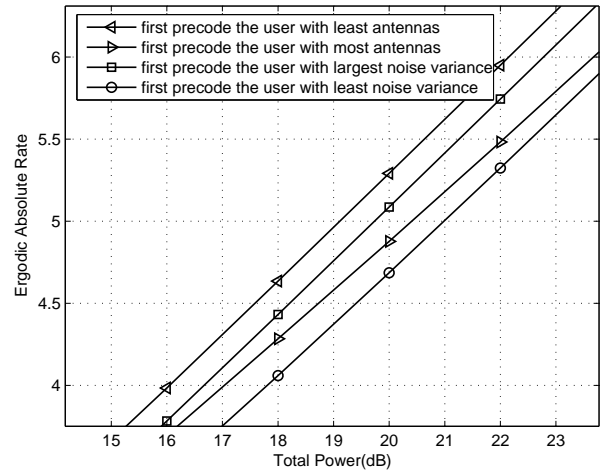}
\caption{The absolute secrecy rate for different users as Bob1
when $J=3$, $K=2$, $N_A=10$, $N_{B1}=1$, $N_{B2}=2$, $N_{B3}=3$,$N_{Ek}=4$.}
\end{figure}


\begin{figure}
  \captionstyle{flushleft}
  \onelinecaptionstrue
  \centering
  \includegraphics[width=3.6in]{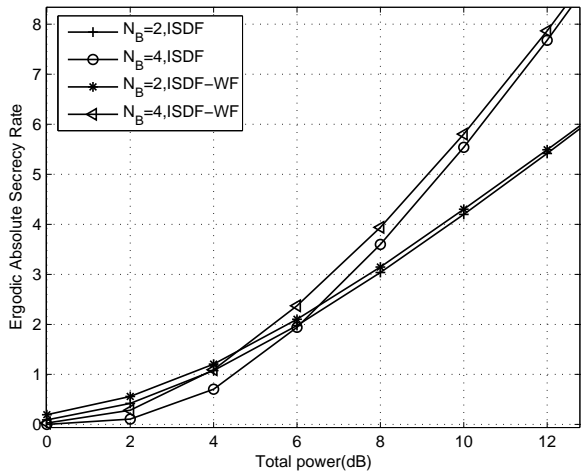}
\caption{Comparison of absolute secrecy rate for ISDF method and
ISDF method with WF when $J=3$, $K=2$, $N_A=10$, $J=3$, $N_{Ek}=4$, $N_{B1}=N_{B2}=N_{B3}=N_{B}$. }
\end{figure}

\subsection{Secrecy Rate and Power Efficiency for Eves' Colluding and Non-Colluding}
Fig.~5 and Fig.~6 show the absolute secrecy rate and information
power ratio of ISDF1 for the Eves' colluding and non-colluding
scenarios. If the Eves choose to wiretap the message jointly, we may
think they are colluding, else non-colluding. As shown in Fig.~5, it
will be more difficult to achieve secure communication if the Eves
choose to cooperate, which is as we expected. We are interesting in
whether we need to allocate more power to transmit information
signal or artificial noise when the Eves choose to cooperate. Fig.~6
shows that more power needs to be allocated to artificial noise if
Eves choose to cooperate. So does the case with more Eves. These demonstrate
that when power has been optimized already and the eavesdroppers' condition
is getting better, the power allocating towards artificial noise can make
more contribution for the secrecy rate than allocating towards users'
information signal.

\subsection{Power Ratio for different Ordering and Power Allocation}
As mentioned in section III-A, since the ordering of the users will
affect the performance,  we wish to study how the power is
allocated between the information signal and the artificial noise
for giving different user priority of precoding. Fig.~7 illustrates
that the the system performs better in term of absolute secrecy
capacity when we give the user with least antennas or largest noise
variance the priority of precoding. This is because that the absolute
secrecy capacity is mainly determined by the poorest-performance
receive-wiretap pair to a large extent. In order to get a large secrecy
rate, we should make the secrecy rate for each user equivalently.
Therefor should give the weaker user (less antennas or larger noise) more
SDF by rendering them the priority of precoding. Fig.~8 shows that the
secrecy capacity can be further increased as introduced in IV-C, if the
water-filling (WF) algorithm is used.

\section{Conclusions}

This paper proposes the precoding strategy based on the ISDF method
for providing secure communication at the physical layer in broadcast
MUME-MIMO wiretap channels combined with artificial noise. We derive
both the secrecy sum rate and absolute secrecy rate for the
proposed ISDF1 and ISDF2 method. Simulations show that the ISDF2
perform best in low SNR region and ISDF1 outperforms other four
methods in high SNR region in terms of achievable secrecy rate.
Furthermore, we find that more power should be allocated to
artificial noise instead of information signal when the
eavesdroppers' condition is better than the intended users, and we
should first precode the user with bad condition (least antennas or
largest noise variance).

\end{document}